\documentclass{elsart}
\usepackage{epsfig}

\def\url#1{#1} 

\begin{document}

\begin{frontmatter}

\title{General considerations on the illumination of galactic nebulae.}

\author{Fr\'ed\'eric \snm Zagury \thanksref{jsps}}

\address{Department of Astrophysics, Nagoya University,
\cty  Nagoya, 464-01 \cny Japan
\thanksref{email} }

\thanks[jsps]{Supported by the Japanese Society
for the Promotion of
Science,  grant ID No P97234.}
   \thanks[email]{E-mail: zagury@a.phys.nagoya-u.ac.jp}

\received{5 June 2000}

 \begin{abstract}
The interpretation of observations over different 
wavelength domains, which now exist over a large fraction 
of the sky, will be used to determine relationships 
between a nebula and its' illuminating source.

The illuminating source of 
a high latitude nebula must be behind the 
cloud and e.g. at close angular distance.
It is either a point source behind the 
nebula or the 
average radiation field created by all the background stars.
In the latter case it is possible to estimate the maximum surface 
brightness a nebula can reach.
Illumination by the galactic plane is always negligible in up to date 
observations.

The red color of some nebulae, when they are not HII regions, is more likely 
to be due to large column densities than to an emission process.

Concerning interstellar grain properties, the same data will be used 
to support the well known property that grains scatter starlight in 
forward direction.

   \end{abstract} 

\end{frontmatter}

 \section{Introduction}
The aim of the paper is to use the complementarity between 
different sky surveys to derive conclusions 
concerning the scattering of starlight in the galaxy.
General and qualitative analysis of the data will lead to
remarkable links between a nebula and its' illuminating 
source.

Section~\ref{data} presents the data to be discussed in 
the paper.
Section~\ref{preview} recalls existing interpretations of
the origin of the light 
scattered by high latitude nebulae, and on the red color of some nebulae.
These interpretations raise questions which will be addressed in 
section~\ref{fs}.
From this section to section~\ref{questions} on, I will arrive at the 
different results of the paper.
Two applications are discussed in sections~\ref{mcldill} and \ref{lmc}.
The results are summarized in the conclusion.
\section{Data}\label{data}
\subsection{Palomar optical sky survey}\label{palomar}
The most representative and complete existing data set of 
interstellar dust 
optical emission is the Palomar plates for the Northern hemisphere and 
ESO plates for the Southern sky. Palomar images in the $B$ and $R$ 
bands 
have a sensitivity of $27\,$Mjy/arcsec$^2$ and a resolution of $2$''. 

Optical interstellar features on the plates have been classified by 
Lynds in 2 
papers, `Catalog of Dark nebulae'~\cite{lynds62} and `Catalog of 
Bright nebulae'~\cite{lynds65}. 

Lynds' dark nebulae (LDN) are highly extinguished areas of sky, many 
of which correspond to molecular clouds (Magnani et al \cite{magnani85}, Lee~\& Myers 
\cite{lee99} and references therein). 
The important extinction of these regions is due 
to the absorption of starlight by interstellar grains. 

LDNs are spread all over the sky, at all galactic latitudes (Lynds 
\cite{lynds62}). The number of LDNs, of course, decreases when moving 
away from the galactic plane. Most LDNs are 
concentrated at $|b|<10$, but many are still observed up to $|b|=25$. 
Lynds' search of LDNs is limited to $|b|<40$.

Bright nebulae are seen in emission on Palomar plates, either in $B$, 
or $R$, or both bands. Like dark nebulae they usually have a small surface coverage 
($<1^{{\circ}\,2}$). Lynds'  bright nebulae (LBN) are
concentrated in the Galactic plane, but are found up to the limit 
of completeness of her catalog, $|b|=40$. 
Lynds' classification of bright nebulae involves various criteria among 
which are color and surface brightness. 
Bright nebulae span a wide range of $B-R$ colors, from completely blue to 
completely red. 
\subsection{Magnani et al. search for high galactic latitude 
molecular 
gas}\label{magnani}
Magnani et al.~\cite{magnani85} made a systematic search for CO in
Palomar extinguished regions at high latitude ($|b|>25$). They detected 
CO in 133 of 488 observed positions. The regions where CO was 
detected were grouped into 35 separate large complexes (MBM 
complexes). The average 
distance to the sun of those complexes was estimated to be $\sim100\,$pc
(Magnani \& de~Vries~\cite{magnani86}). 
Most of MBM 
complexes have some Lynds bright nebula 
associated with them: six blue nebulae, ten 
deemed of equal brightness in both $R$ and $B$ plates, four brighter 
on red plates and four visible on red plates only.  Remarkably, 
the CO emission does not systematically coincide with the LBNs. 
In other words, there is dense material which does not scatter starlight  at optical 
wavelengths at a level detectable by the Palomar survey.

\subsection{MCLD123.5+24.9}\label{mcld}
I will also mention the optical images obtained at Kitt Peak 
Observatory by Roc Cutri and Fr\'ed\'eric Zagury. The I image of the 
$1^{\circ}\times1^{\circ}$ field is presented here 
(figure~\ref{poli}). 
This field can be perceived on Palomar plates, but with far less 
detail than in our images. Zagury, Boulanger \& 
Banchet~\cite{zagury99} (hereafter paper~I) 
have shown that, despite the presence of local red filaments across the field, all the 
optical emission of MCLD123.5+24.9 is well explained by normal scattering 
processes and illumination by the North star, $1^{\circ}$ away. 

Falgarone et al.~\cite{falgarone} have mapped in molecular lines the 
area inside the rectangle shown in figure~\ref{poli}. Their study reveals 
high densities,  $n_{H_2}\gg$ a few 
$10^3\,\rm cm^{-3}$, and column densities in the feature of high extinction 
within the rectangle.
Densities increase toward the central and most extinguished parts. 
The dense core is completely extinguished in the $B$ 
band (figure~2 of paper~I) and has a definite red color since 
its' densest part remain bright in the $I$-band and, but to a 
lesser extent, in the $R$-band.

MCLD123.5+24.9 is centered at $l=123.5$ and $b=24.9$. 
It is estimated to be $\sim 120\,$pc to the sun (paper~I), which 
is in accordance with the average distance estimate of  high latitude clouds 
due to Magnani~\& de~Vries~\cite{magnani86}.

\subsection{IRAS whole sky survey}\label{iras}
\begin{figure}
\resizebox{0.8\columnwidth }{!}{\includegraphics{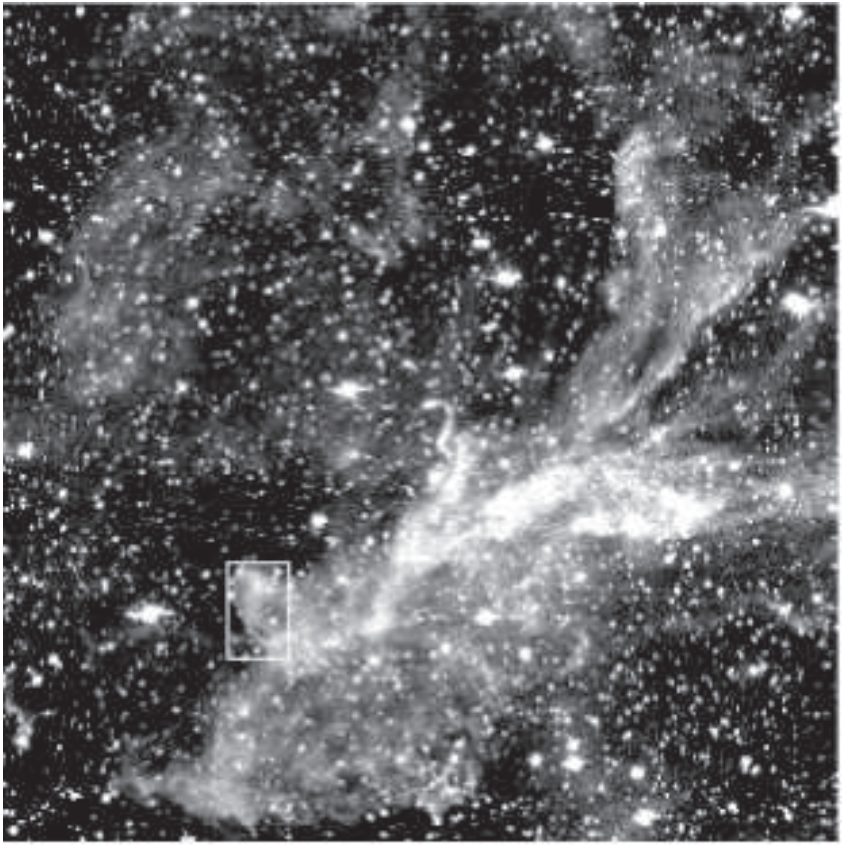}} 
\caption{$I$  band image of MCLD123.5+24.9. The resolution is $2$''. 
The field is $\sim 1^{\circ}\times 1^{\circ}$. 
North is at the upper edge and East at the left edge of the 
image. Polaris is $1^{\circ}$ North to the field. The extinct region 
seen in 
absorption, and often mentioned in the text is enclosed in the 
rectangle. 
This image is discussed in Zagury, Boulanger and Banchet~\cite{zagury99}.} 
\label{poli}
\end{figure}

The IRAS sky survey has 
revealed large scale structures which HLNs and MBM complexes 
are part of.
It has also evidenced the complex structure of galactic cirrus. 
This structure is also perceived at a scale $100$ times smaller than reached 
by IRAS in the optical emission of MCLD123.5+24.9 (figure~\ref{poli}).

The $100\,\mu$m emission of interstellar matter is the thermal 
emission of the same large grains which scatter starlight in the 
optical.
When a cloud is illuminated by a source of light, there is a very 
good similarity between the optical image of the cloud and the IRAS 
$100\,\mu$m image of its' infrared emission (figure~2 of paper~I).
\section{A review of current hypothesis concerning the color and 
 illumination of HLNs}\label{preview}
Many studies of nebulae have based their analysis 
on the comparison of the color of the nebulae and that of the 
illuminating source.
Lynds uses the difference in color between nebulae to 
distinguish `reflection nebulae', `purely emitting regions' and 
`transition nebulae', 
which conforms to the idea that scattering should turn the emission
to blue, because dust scatters blue light more efficiently than 
red light. According to Lynds, the `blue'
emission of reflection nebulae should be attributed to the scattering 
of starlight, while the red regions arise from some emission line. 
The same idea is taken up 
and furthered in more recent studies. 
It leads to the notion of Extended Red Emission (ERE) 
(see Gordon, Witt and Friedmann \cite{gordon} 
and Leinert et al. \cite{leinert98} for a review). 

The color of a nebula depends on its' optical depth and on the color 
of the illuminating radiation field. 
Of course, the latter needs to be unambiguously identified.
The problem of the illuminating source of the blue high latitude nebulae 
was first quantitatively addressed by Sandage~\cite{sandage}.
In this paper Sandage shows that a high latitude cloud of low column density 
in which grains isotropically scatter starlight will be seen on the 
Palomar plates with a surface brightness of magnitude close to the 
magnitude observed for the HLNs.
Like it was already suggested by Lynds~\cite{lynds65}, Sandage~\cite{sandage} and 
Magnani et al.~\cite{magnani85} attribute the blue HLNs of the Palomar plates to 
dusty nebulae reflecting the light of the galactic plane.

Stark~\cite{stark} tried to improve 
Sandage's model and included forward scattering in his Monte Carlo 
simulations.  
A look at figure~9 of Stark's paper shows that the HLNs surface brightness would 
increase the surface brightness given in Sandage by $\rm 
3\,mag/arcsec^2$, making HLNs undetectable at Palomar sensitivity. 

Some authors (Guhathakurta~\& Tyson~\cite{gu1}, Guhathakurta~\& Cutri~\cite{gu2}) 
have used Sandage's proposal as a basis for their 
affirmation that red high latitude nebulosities `produce' ERE.
According to these authors, high latitude clouds are low column 
density clouds reflecting the light of the galactic plane.
With such an illuminating source, the observed color of the clouds is too red to 
be explained by scattering only. 
It implies an emission process which occurs within the clouds from the 
red to the near infrared.
\section{The illumination source of high latitude nebulae.} \label{fs}
If HLNs are indeed reflecting the light from 
the galactic plane, which is `everywhere, all the time' (Sandage \cite{sandage}), 
how is it that the light scattered by most dense clouds detected in 
CO by Magnani et al.~\cite{magnani85}
is not observed on the Palomar plates? 
Why is the high density and high
column density feature of figure~\ref{poli} seen in absorption in the 
$B$ band? 

At the time of Sandage's paper IRAS had not yet been launched and 
HLNs could be thought to represent a large part of existing interstellar high 
latitude clouds. 
Since the IRAS survey, we know that HLNs are a very small 
fraction of infrared cirrus. 
If HLNs reflect the Galactic 
plane's light, we expect a much larger fraction of cirrus clouds 
to be bright in the $B$ and/or $R$ bands.
Why do HLNs represent only a small part of the cirrus?

These questions call for several conclusions.

A first conclusion is that HLNs are never illuminated by the galactic plane.
Most HLCs are in the solar neighborhood.
Viewed from earth, they occupy comparable positions 
in regard of the light they reflect from the galactic plane.
If the galactic plane provides the illumination of one high latitude 
nebulosity it will, as noted by Sandage~\cite{sandage}, similarly 
illuminate all high latitude clouds and explain the surface brightness of all HLNs.
Hence, if a low column density medium reflects the light from the Galactic plane
at a level detectable to Palomar sensitivity, a much larger fraction of 
the HLCs should appear on the Palomar plates.
This already indicates that the galactic plane is unlikely to be the 
illuminating source of HLNs.

Furthermore, if the galactic plane is the illuminating source of low column 
density HLNs, higher surface brightness is also expected from the reflection 
by the high column density parts of the infrared cirrus. 
Since  high latitude dark nebulae above the galactic plane are 
completely extinguished (LDNs, MBM dark clouds, the dark feature in 
MCLD123.5+24.9), the source of light of HLNs cannot be on the same side as the 
observer. It has to be behind the nebulae.

Secondly, interstellar grains must strongly scatter light in forward 
direction.
This is of course known since Henyey \& Greenstein~\cite{henyey}.
It can also be uncovered from Sandage's calculations since, in case of
near isotropic scattering, high 
latitude cirrus would be detectable at Palomar sensitivity.
We can assume there is an angle of scattering $\varphi_f\ll \pi /2$ 
within which most photons are certainly deviated from their initial 
trajectory, after being scattered 
by interstellar grains.

Hence, the illuminating source of an HLN must be 
behind and, because of forward scattering, at close angular distance from the nebula.
This is confirmed by the observations: 
all bright nebulae mentioned in literature are at most a few 
tens of arcminute from their illuminating star.

It is important to remark that the low column density 
approximation which is usually made to estimate the surface brightness 
of a nebula is probably not suitable to describe bright nebulae.
Contrary to this hypothesis, the most general case shoud be that a cloud 
is visible because it has sufficient column density to scatter enough starlight to 
be detected: its optical depth should in general be close to $1$.
This is the case for MCLD123.5+24.9 where too low column density regions have 
little emission and where high density clumps are seen in absorption.
In between, regions for which the optical depth gives maximum surface 
brightness are luminous on the image (paper~I).
These regions are not the same for the $B$-band than for the $R$-band 
or for the $I$-band.
Since cirrus are known for their small scale structure and for 
experiencing large ranges of column densities, the same should 
generally be held for all cirrus.

Different types of background illuminating sources can be thought of 
for the illumination of an HLN. 
They are reviewed in the following section.

\section{The possible illuminating sources of nebulae} \label{cas}
As a consequence of section~\ref{fs}, high latitude nebulae seen in
emission (HLNs) should be illuminated by a background source.  The
source radiation field can be of 3 different types:
\begin{enumerate}
\item
\label{cas1} 
A radiation field whose source of light comes from one or a few 
background point sources (e.g. bright 
stars). The number of possible illuminating sources of a cloud  
decreases when moving away from the Galactic 
plane which renders the identification easier.  
 \item
 \label{cas2}
Multi-directional light sources, none of which dominate the others. 
This is the case where all background stars
 contribute to the cloud illumination. For purposes of calculation, 
the 
 radiation field in one direction can be approximated by an average 
 intensity per unit solid angle ($\sigma$).
 \item
 \label{cas3} 
 The radiation field is isotropic 
 (at least for directions within $\varphi_{f}$). 
 \end{enumerate}
Case~\ref{cas3} cannot produce any optical emission. 
Each scattered photon from a direction $\varphi_1$ to a 
direction $\varphi_2$ will be 
replaced by a photon scattered from  $\varphi_2$ to $\varphi_1$. 
Pixels looking outside the cloud will receive more 
photons than pixels inside the cloud due to absorption. The cloud will 
be seen in absorption. Whatever the 
radiation field is at the cloud location, only that part which presents 
some kind of anisotropy should be taken into account to explain the optical 
emission of an interstellar cloud. 
In sections~\ref{1source} and \ref{sbnbsource}, cases~\ref{cas1} and
\ref{cas2} will be investigated in further detail.
   \section{Nebulae illuminated by a background point source}
   \label{1source} 
In general, for nebulae illuminated by a star, different 
data sets in the optical wavelength and in the UV wavelength ranges 
indicate a power law decrease of the surface 
brightness with angular distance to the illuminating star.
Witt \cite{witt80,witt92}) finds an exponent of $1.5$.
Zagury~\cite{zagury00a} finds $2$.  As it is shown in the latter
paper, the quick decrease of the surface brightness with angular
distance to the star confirms a strong forward scattering phase
function.  The power law dependence of the surface brightness on
angular distance to the star also shows that the illuminating source
must be close to the nebula, as noted in section~\ref{fs}.

Let $\varphi$ be the angle of scattering, $\theta$ the angular 
distance of a nebula from its' illuminating star, $d$ and $D$ the 
respective distances of the star to the nebula and to the observer.
These notations correspond to the figure~8 of paper~I.
We have $\varphi=(D/d)\theta$ and $\varphi>\theta$ since $D>d$.  
From figure~8 of paper~I, it is clear that, if $\theta$ and the 
distance earth-nebula are kept 
constant, $\varphi$ decreases when $D$ increases.  
Because of forward scattering, for a given and constant strength of the
radiation field at the cloud location, the nebula surface brightness
will increase when the star is moved away.  
Hence, for a given
sensitivity of the observation and for a constant value of the
radiation field at the cloud location, the angular distance $\theta$ from the
source at which a nebula can be detected increases when the star is
far away.  The maximum angular distance at which the nebula can be
observed with a given sensitivity and strength of the radiation field
at the cloud location is obtained for a star at infinity.  In this
case $\theta=\varphi$ and $D/d=1$.

It was noted in paper~I that Polaris may illuminate MCLD123.5+24.9 and have a
negligible contribution to cloud heating.  
This is possible because the star will have its' light efficiently scattered
due to the small angle of scattering while the heating of the dust
by Polaris can be small compared to the radiation field due to all the stars. 
Geometry can favor scattering of the light of a particular star
while heating is sensitive to the integrated radiation field.
   \section{The surface brightness of a nebula illuminated by a large number of point sources } 
   \label{sbnbsource}
The illuminating source is supposed to be all the background stars. 
Because of forward scattering stars within a certain angle 
$\theta_{max}$ from the direction of the nebula will make 
a major contribution to the cloud surface brightness. 
Pixels looking outside the cloud and in between $2$ 
stars will receive no power, while pixels looking at the cloud will 
receive the light diffused by grains inside the cirrus.

If stars are evenly distributed in the sky, 
at least within $\theta_{max}$ from the direction of the cloud,
photons scattered once will 
create an isotropic radiation field (within $\theta_{max}$), so that 
further scatterings will not increase the cloud surface brightness 
(case~\ref{cas3} of section~\ref{cas}). The single scattering approximation
proposed in paper~I can be applied. With an average illumination by unit 
solid angle 
$\sigma$, due to stars within $\theta_{max}$, in the cloud vicinity, 
the cirrus surface brightness will 
remain proportional to $\sigma$ and will not depend on the phase 
function:
\begin{eqnarray}
S&=&\sum_{stars 
\,\star}(g(\varphi_\star)\omega{\tau}e^{-\tau}F_\star) \\
&=&\sigma\omega{\tau}e^{-\tau}\int_{\varphi}2{\pi}g(\varphi){\rm d}\varphi\\
&=&\sigma\omega{\tau}e^{-\tau}
\label{eq:sigma}  
\end{eqnarray}
The surface brightness reaches its maximum for $\tau=1$:
\begin{equation}
S_{max}\,=\,0.37\omega\sigma\sim 0.2\sigma,
\label{eq:sbsigma}  
\end{equation}
with $\omega=0.6$. 
If units of mag per unit solid angle are adopted:
\begin{equation}
S_{max}\,=\,m_\sigma+1.7,
\label{eq:sbmsigma}  
\end{equation}
with $m_\sigma$ the surface brightness of the background sources.

High latitude cirrus are close to the earth, compared to the stars' 
scale--height ($\sim$~$400$~pc). 
For the purpose of calculation, we can suppose that the radiation field 
seen by an HLC is the same we see 
on earth and estimate $\sigma$ from galactic models such as the 
Besan\c con Galactic model ().
   \section{An example: the illumination source of MCLD123.5+24.9 } 
   \label{mcldill}
In paper~I, illumination of MCLD123.5+24.9 was studied under the assumption 
that the field is illuminated by Polaris. The maximum surface 
brightness over the field, above the absorbed region mapped by 
Falgarone et al.~\cite{falgarone}, $\sim0.032\pm0.003$MJy/sr 
($24\,$mag/arcsec$^2$) in the $R$ 
band, can be explained by illumination by Polaris, even if the star 
does not contribute to the Polaris Flare heating. 
 Paper~I bases itself on the hypothesis that Polaris is behind 
the cirrus. This hypothesis, although credible, has not yet been proven. If 
Polaris was in front of the Polaris Flare, illumination of MCLD123.5+24.9 by the star 
would probably have to be reviewed. Another explanation, for instance 
illumination by all the background stars, will need to be considered.

From the
Guide Star Catalog and Besan\c con Galactic model, assuming the cirrus 
is illuminated by the same 
distribution of stars we see from Earth, $\rm \sigma_R$ is found to be
$\sigma_R\sim 0.1\,$MJy/sr or 
$\sim23\,$mag/arcsec$^2$ in MCLD123.5+24.9 direction. 
If the field was illuminated by all the background stars, 
the expected maximum $R$ surface brightness over the field would be 
$0.02\,$MJy/sr or $24.5\,$mag/arcsec$^2$ 
(equation~\ref{eq:sbsigma}), close to what is observed.  The surface
brightness of the very low column density medium, of average
extinction $\rm A_V\sim 0.5$ ($\rm A_R\sim 0.37$), in which
MCLD123.5+24.9 seems to be embedded (see paper~I), will be
$\sigma{\omega}{\tau}e^{-\tau}=0.015\,$MJy/sr
(equation~\ref{eq:sigma}).  The difference between the 2 values,
$0.005\,$MJy/sr, is the quantity which should be compared to the
$0.032\,$MJy/sr found on the $R$ image.  If this interpretation is
true, background stars do not provide the luminosity to account for
MCLD123.5+24.9 optical emission.

For constant albedo and phase function, the maximum brightnesses over the field 
in different bands, assuming that
they are reached at some (not necessarily the same for each band)  
pixel in the field, should scale like the source radiation field 
intensities. In MCLD123.5+24.9's case: 
$SB_{B,max}/SB_{I,max}\sim (0.6\pm0.1)$ and 
$SB_{R,max}/SB_{I,max}\sim1$. The colors of Polaris are
$F^0_{B}/F^0_{I}\sim0.5-0.7$, 
$F^0_{R}/F^0_{I}\sim1$ (paper~I, table 1), in accordance 
with the previous values. For the background sources model, 
we have $\sigma_{B,max}/\sigma_{I,max}\sim 0.5$, 
$\sigma_{R,max}/\sigma_{I,max}\sim 0.89$. If stars with magnitudes of less 
than 6 were included, both colors would be diminished.

This argumentation is nevertheless model dependent. I do not feel 
it is conclusive so the choice remains between the cirrus being in 
front of Polaris and illuminated by it, or being behind Polaris and 
illuminated by all the background stars. In the latter case its distance to the 
sun should be greater than $130\,$pc. More definitive conclusions
should come from the comparison of spectroscopic observations of MCLD123.5+24.9 
and Polaris or by precise determination of the cloud distance from the sun. 
   \section{Second example: a cirrus illuminated by the LMC? } 
   \label{lmc}
\begin{figure*}
\resizebox{0.9\columnwidth }{!}{\includegraphics{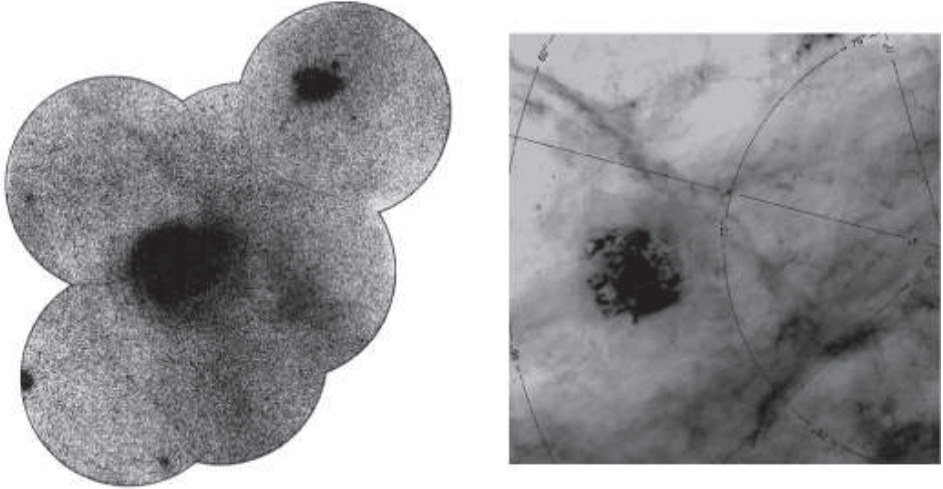}} 
\caption{De Vaucouleurs'~\cite{devau} image of the Magellanic Clouds is on the left. The filament 
can be seen crossing the field above the LMC. The right image is 
IRAS $100\mu$m image of the region. The filament is part of a larger Galactic cirrus.} 
\label{imdevau}
\end{figure*}
In 1955,  G. de Vaucouleurs (de Vaucouleurs~\cite{devau}) 
published a study of the LMC. Plate II and figure 2 of de Vaucouleurs' paper 
show a long filament extending $10^{\circ}$ far, West and North, from the LMC. 
The filament surface brightness ranges between $25\,$mag/arcsec$^2$ and 
$27\,$mag/arcsec$^2$ at its most distant part. 
Figure~\ref{imdevau} reproduces de Vaucouleurs image along with the IRAS 
$100\,\mu$m image of the same region.

Originally (de Vaucouleurs~\cite{devau}) this feature was interpreted as a 
possible spiral arm of the LMC, though later detailed searches 
revealed a puzzling absence of stars (W.~Kunkel, private 
communication). Comparison with the IRAS 
$100\,\mu$m image (figure~\ref{imdevau}) shows it is in fact part of 
a large high latitude galactic cirrus. 
Its distance to the sun is estimated to be between $70\,$pc 
(Wang \& Yu~\cite{wang})
and $200\,$pc (Penprase et al.~\cite{penprase}). 

The IRAS
$100\,\mu$m surface brightness of the filament does not exceed 
$6\,$MJy/sr and decreases toward its northern extremity. 
This decrease can be interpreted as a decrease of column density along 
the line of sight.
It may also be attributed to the attenuation of the heating because it
coincides with the decrease of the ambient ISRF:
the filament extends toward high galactic latitudes, from $b=-33$ to $b=-45$.

Two radiation 
fields can be thought of to explain the filament optical brightness.

One is background stars.
From the Besancon model of our galaxy, the surface brightness of the background stars
at optical wavelengths is 
of order $30\,\rm mag/arcsec^2$ at the LMC position, and 
decreases with increasing latitude.
According to equation~\ref{eq:sbmsigma}, scattering of the background 
starlight by dust in the cirrus will give a surface brightness of 
$30\,\rm mag/arcsec^2$ or more.  
If so, background stars are not luminous
enough to account for the cirrus optical brightness.

The only luminous object of the region is the LMC.
The LMC has an apparent 
magnitude $B_0^T\sim 0.9\,$mag and a surface brightness of $m'_{25}\sim 22\,\rm mag/arcsec^2$
(Third Reference Catalogue of Bright Galaxies, de Vaucouleurs 1991,  ).
Correction for  extinction by the cirrus will decrease these values. 
In the LMC immediat vicinity, equation~\ref{eq:sbmsigma} shows that
illumination by the LMC will explain the filament brightness. 
Assuming a decrease of the brightness as $\theta^{-2}$ at most,
section~\ref{1source}, the expected decrease of the surface brightness
of the filament over $10^\circ$ (corresponding to the extent of the 
filament) will be at most
$5\,\rm mag/arcsec^2$, which is compatible with the observations.  The
large area over which the LMC filament was observed, much larger than 
observed for nebulae illuminated by a close star, agrees with the discussion of
section~\ref{1source} on the angular extent up to which the scattered
emission of a nebula is detectable.
   \section{Illumination of high latitude clouds by the ISRF } \label{dglillu}
The ISRF from one direction comprises the light from the stars in 
that direction and the diffuse galactic light (DGL) which is the 
starlight scattered by interstellar grains.
As indicated in a previous paper (Zagury, `Is there ERE at high 
galactic latitude?', submitted), and because grains forward scatter light, DGL in 
one direction can logically be attributed to be the light of background stars scattered by dust embedded 
in the nearby cirrus on the line of sight.

Contrary to the direct starlight, DGL cannot contribute to the 
illumination of a cirrus.
This is a straightforward consequence of section~\ref{sbnbsource} 
(second paragraph).
The maximum surface brightness of a cirrus due to the illumination
by background stars is given by equation~\ref{eq:sbsigma} of
section~\ref{sbnbsource}: if there is no bright star in the vicinity of
the cirrus its' surface brightness is at most $20 \% $ of the
background stars surface brightness.  Since the surface brightness of
background stars and galaxies decreases with increasing latitude, we
can expect an average decrease of the cirrus' surface brightness with
increasing latitude.  Such a decrease is observed, as shown by the
studies of Toller~\cite{toller81} and of Gordon et al.~\cite{gordon}.  
As expected it is proportional to the decrease of starlight.

Concerning the specific case of the high latitude cirrus detectable on 
the Palomar plates, with a brightness level above average,
the distribution of Palomar HLNs all over the sky, 
with no evident correlation between surface 
brightness and galactic latitude, along with the much higher optical 
emission of HLNs, may require the presence of a 
luminous object close to the cloud, as suggested 
for MCLD123.5+24.9 (paper~I and section~\ref{mcld}) and for the LMC
filament (section~\ref{lmc}).

\section{The red color of the nebulae}\label{questions}
Why should red areas in the sky act as emitting regions? 
After all, the sky (on earth) is blue in most cases, but 
the horizon may be red at sunset or sunrise. 
No one will ever claim 
there are crystalline nano particles of pure silicon (Witt~\cite{witt98}) 
or any other kind of emitting particles which
luminesce at those particular times of the day.  The red color of
the sky comes from the increase of column density crossed by light
when the sun rises or descends at the horizon.  Why shouldn't Lynds' red
nebulae, similarly and unless they are proved to be HII regions, be
the consequence of variations in column density?

The $10'\times 10'$ red feature of MCLD123.5+24.9, mapped by Falgarone 
et al.~\cite{falgarone} 
in molecular lines is seen in absorption and totally extinguished on the blue image 
(section~\ref{mcld} and paper~I). 
The high column densities and densities found in this 
area confirm the link between red nebulae and high column density regions.
In paper~I we have explained the color variations of the nebula by 
variations of the column density and normal scattering properties of 
the grains.
We have shown that there was no need for an additional component of 
red light created within the nebula.

The excess of red emission Guhathakurta~\& Tyson~\cite{gu1} and Guhathakurta~\& 
Cutri~\cite{gu2} found for nebulae similar to MCLD123.5+24.9 relies on a 
comparison of the nebulae color and the color of the galactic plane.
It assumes the nebulae are 
illuminated by the galactic plane, which is difficult to accredit
(section~\ref{fs}).  The conclusion of both papers, that an additional
ERE emission was necessary to justify the color of the nebulae, is now
questionable.  The Guhathakurta~\& Tyson~\cite{gu1} paper further
assumes isotropic scattering to discard nearby stars as possible
illuminating sources of the nebulae.  
Forward scattering will potentially allow at least 
one star for the illumination of each of the four fields 
(see figure~7 in Guhathakurta~\& Tyson~\cite{gu1}).
This will of course completely modify their analysis as well as their
conclusions.  It follows, and unless more precise studies will prove
the contrary, that no up to date observation implies another process
than scattering to explain the color of high latitude clouds.

\section{Conclusion} \label{con}
Complementary sets of data were compared to 
obtain a better understanding of the relation between nebulae and their
illuminating source.  The paper has focused on high latitude nebulae
for which the source of light is more easily identified, but there is
little doubt that the results gained for high latitude clouds still
apply closer to the galactic plane.

The existence of a large number of high column density regions which 
are extinguished on the Palomar plates shows that the contribution of the galactic plane 
to the illumination of the infrared cirrus is negligible.
It also implies that interstellar grains scatter light in forward 
direction, a property first uncovered by Henyey~\& Greenstein~\cite{henyey}.

The illuminating source of a high latitude cloud must be behind the
cloud and at close angular distance from it.  
It must present some anisotropy.  
It can be either the background stars in the direction of
the cloud or a nearby star which dominates the radiation field over 
the directions close to the cloud line of sight.  

As noted in paper~I, a particular star can be responsible for the cloud
optical emission without giving significant contribution to the
heating of the cloud.
This is because background stars close to the cloud line of sight have
their light efficiently scattered in the direction of the observer
while stars from all directions contribute to the heating of the
cloud.
For a given intensity of the radiation field
at the cloud position, the surface brightness of the cloud increases with the
distance of the star from the cloud.  
This property may explain the large extent of a filament close to 
the LMC, detected at optical wavelengths by de~Vaucouleurs~\cite{devau}.
Both its' surface brightness and the spatial extent of the emission would agree 
with illumination by the LMC.

When the scattered light cannot be attributed to an individual (eventually a few) star 
the surface brightness of the cloud will be at most
$20 \%$ of the surface brightness of the background stars.  The
radiation field created by the scattering of the light from the background stars
can not produce illumination of another cloud: DGL will not
induce an increase of the surface brightness of high latitude clouds.

A precise determination of the illuminating source of a nebula is 
necessary before reaching any conclusion concerning its' color.
The red color of some nebulae can be explained by higher column densities 
than in blue nebulae.
Up to date observations do not seem to be able to prove the existence 
of any other process than scattering to explain the color of high 
latitude nebulae.

{}

\end{document}